\documentclass[twocolumn,showpacs,preprintnumbers,amsmath,amssymb]{revtex4}

\usepackage{graphicx}
\usepackage{dcolumn}
\usepackage{bm}

\begin{document}


\title{Shape and structure of N=Z $^{64}$Ge; Electromagnetic transition rates 
from the application of the Recoil Distance Method to knock-out reaction.}

\author{K.~Starosta$^{1,2}$}
\author{A.~Dewald$^3$}
\author{A.~Dunomes$^2$}
\author{P.~Adrich$^1$}
\author{A.M.~Amthor$^{1,2}$}
\author{T.~Baumann$^1$}
\author{D.~Bazin$^1$}
\author{M.~Bowen$^{1,2}$}
\author{B.A.~Brown$^{1,2}$}
\author{A.~Chester$^{1,2}$}
\author{A.~Gade$^{1,2}$}
\author{D.~Galaviz$^1$}
\author{T.~Glasmacher$^{1,2}$}
\author{T.~Ginter$^1$}
\author{M.~Hausmann$^1$}
\author{M.~Horoi$^4$}
\author{J.~Jolie$^3$}
\author{B.~Melon$^3$}
\author{D.~Miller$^{1,2}$}
\author{V.~Moeller$^{1,2}$}
\author{R. P. Norris$^{1,2}$}
\author{T. Pissulla$^{3}$}
\author{M.~Portillo$^1$}
\author{W.~Rother$^3$}
\author{Y.~Shimbara$^1$}
\author{A.~Stolz$^1$}
\author{C.~Vaman$^1$}
\author{P.~Voss$^{1,2}$}
\author{D. Weisshaar$^1$}
\author{V. Zelevinsky$^{1,2}$}

\affiliation{ 
$^1$National Superconducting Cyclotron Laboratory, Michigan State
University, East Lansing, Michigan 48824, USA\\
$^2$Department of Physics and Astronomy, Michigan State University,
East Lansing, Michigan 48824, USA\\
$^3$ Institut f{\"u}r Kernphysik der Universt{\"a}t zu K{\"o}ln, D-50937
K{\"o}ln, Germany\\
$^4$ Department of Physics, Central Michigan University, Mount
Pleasant, Michigan 48859, USA
}

\date{\today}

\begin{abstract}

Transition rate measurements are reported for the $2^+_1$ and $2^+_2$
states in N=Z $^{64}$Ge. The experimental results are in excellent
agreement with large-scale Shell Model calculations applying the
recently developed GXPF1A interactions. Theoretical analysis suggests
that $^{64}$Ge is a collective $\gamma$-soft anharmonic vibrator. The
measurement was done using the Recoil Distance Method (RDM) and a
unique combination of state-of-the-art instruments at the National
Superconducting Cyclotron Laboratory (NSCL). States of interest were
populated via an intermediate-energy single-neutron knock-out
reaction. RDM studies of knock-out and fragmentation reaction products
hold the promise of reaching far from stability and providing lifetime
information for excited states in a wide range of nuclei.

\end{abstract}
\maketitle

Experiments involving N=Z nuclei play a vital role in the
understanding of nuclear structure.  Along the N=Z line, protons and
neutrons occupy the same Shell-Model orbitals. The resulting overlap
of nucleon wave functions leads to an amplification of the residual
proton-neutron interactions. In nuclei with 28 $<$ N = Z $<$ 50, large
shell gaps open simultaneously for prolate and oblate quadrupole
deformations. Atomic nuclei in this region are a subject of vigorous
experimental studies due to a remarkable diversity of
shapes. Variations in the excitation energy of low-lying states in
this region are often used to analyze the evolution of structure away
from the doubly-magic $^{56}$Ni core. However, electromagnetic
transition rates are recognized as providing a more sensitive probe of
collectivity and deformation. The current letter reports on the
application of the Recoil Distance Method (RDM)\ \cite{Che06} to
lifetime studies of N=Z=32 $^{64}$Ge.

Successful application of the RDM opens up new possibilities for
lifetime measurements of excited nuclear states at fragmentation
facilities. In the experiment reported here, states of interest were
populated via an intermediate-energy single-neutron knock-out from
rare isotope beams of $^{65}$Ge and $^{63}$Zn. The measurement took
advantage of state-of-the-art instruments available at the NSCL; it
brought together the Coupled Cyclotron Facility\ \cite{mil01} for
acceleration of the primary beams, the A1900 mass separator for rare
isotope selection\ \cite{mor03}, the diamond timing detector for
particle identification of the incoming beam\ \cite{sto06}, the
Segmented Germanium Array (SeGA) for $\gamma$-ray detection\
\cite{mue01}, the K{\"o}ln/NSCL plunger device for the RDM\
\cite{dew06}, and the high-resolution S800 spectrograph for
identification of the reaction products\ \cite{baz03}. This unique
combination offers access to a wide range of exotic nuclei which can
be investigated via the RDM for lifetime information.  Such studies
can provide information on transition rates far from the line of
stability.

Ground-state shapes in 28 $<$ N = Z $<$ 50 nuclei are predicted to
evolve from spherical to triaxial, oblate, prolate and back to
spherical as mass increases\ \cite{kan04} due to occupation of
identical deformation-driving orbitals. Moreover, excited levels can
have significantly different structure than the ground state. For
example, in $^{68}$Se the oblate-deformed ground state band is
reported to coexist with a prolate-deformed excited band \
\cite{fis99}. Currently, the B($E2$,$2^+_1 \rightarrow 0^+_1$) in
even-even N=Z nuclei beyond doubly-magic $^{56}$Ni are known only in
$^{72}$Kr from a recent Coulomb excitation experiment\ \cite{gad05}.
While the single-step Coulomb excitation process is used very
effectively to investigate transition rates to the first excited
state\ \cite{coo06}, the RDM combined with knock-out or fragmentation
reactions provides an opportunity to access states beyond the
$2^+_1$. In particular, the lifetime of the $2^+_1$ and $2^+_2$ states
in $^{64}$Ge are reported in the current study.

\begin{figure*}
\includegraphics[angle=-90,width=18 cm]{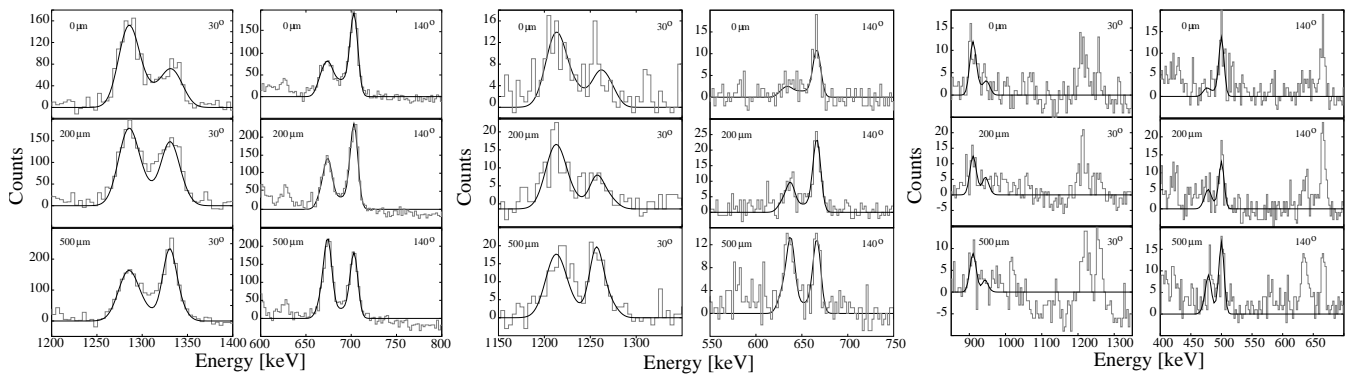}
\caption{\label{fig1} Experimental data and lineshape fits for the
954-keV $2^+_1\rightarrow 0^+_1$ transition in $^{62}$Zn (left),
901-keV $2^+_1\rightarrow 0^+_1$ transition in $^{64}$Ge (middle), and
677-keV $2^+_2\rightarrow 2^+_1$ transition in $^{64}$Ge (right). In
each panel the left/right spectra are for the SeGA rings at
30$^\circ$/140$^\circ$, while top/middle/bottom spectra are for
0/200/500 $\mu$m target/degrader separation, respectively. The Doppler
corrections account for the detector segmentation but not for the
angular position of the detector in the array. See text for further
details on lineshape calculations.}
\end{figure*}

The reduced $E2$ transition rates measured here for the $2^+_1$ and
$2^+_2$ states in $^{64}$Ge are in excellent agreement with the
large-scale Shell Model calculations which use the recently developed
GXPF1A effective Hamiltonian \cite{gx1a}.  GXPF1A was first exploited
in a comprehensive study of energy levels in $^{56}$Ni \cite{ni56}.
GXPF1A was derived from a microscopic calculation by Hjorth-Jensen
based on renormalized $G$-matrix theory with the Bonn-C interaction
\cite{rg}, and was refined by a systematic fitting of the important
linear combinations of two-body matrix elements to low-lying states in
nuclei from $A=47$ to $A=66$, including some states of $^{56}$Ni
\cite{gx1a,gx1}.  The GXPF1A results yield spectroscopic quadrupole
moments of nearly equal magnitude but opposite sign for the $2^+_1$
and $2^+_2$ states in $^{64}$Ge which can be understood from an
anharmonicity in collective vibrations.

In the current experiment the RDM developed for intermediate-energy
Coulomb excitation\ \cite{Che06} was successfully applied in a
measurement of nuclear states populated in single neutron knock-out
reactions. A cocktail beam of rare isotopes comprised of 5\%
$^{65}$Ge, 35\% $^{64}$Ga, 52\% $^{63}$Zn and 8\% $^{62}$Cu was
produced via in-flight projectile fragmentation of $^{78}$Kr at 150
MeV/u as described in\ \cite{sto05}. The constituents of the incoming
beam were identified on an event-by-event basis from the RF time of
flight between the K1200 cyclotron and the timing diamond detector\
\cite{sto06} in the object of the S800 spectrograph\ \cite{baz03}. The
use of the radiation hard diamond for particle identification was
crucial to handle the $\sim$10$^6$ particle-per-second rate of the
incoming beam. The quality of the identification was sufficient to
completely separate incoming beam components in the off-line analysis.

The nuclei of interest for the current study were produced in nuclear
reactions at the target/degrader position of the K{\"o}ln/NSCL plunger
device\ \cite{dew06}. The plunger device was mounted at the target
position of the S800 spectrograph\ \cite{baz03}.  The mass and charge
of the reaction products were extracted on an event-by-event basis
from the time-of-flight and energy-loss information. The time of
flight was measured between the diamond detector in the object of the
S800 and the E1 plastic scintillator in the S800 focal plane. The
energy loss measurement was performed in the ionization chamber at the
S800 focal plane\ \cite{you99}. In the off-line analysis the outgoing
reaction products were identified separately for each component of the
incoming cocktail beam. Below, two channels are discussed;
the single neutron knock-out from $^{63}$Zn leading to $^{62}$Zn and
from $^{65}$Ge leading to $^{64}$Ge. The transition rates in $^{62}$Zn
are known from measurements in stable beam facilities\ \cite{jun00}
and serve here as a consistency check.

The reaction products emerged from the 500~$\mu$m thick $^{nat}$C
plunger target with a velocity of $\beta_H\sim$0.39. Nuclei in excited
states decayed in flight by $\gamma$-ray emission after a distance
governed by the lifetime of the state.  A stationary 250~$\mu$m thick
$^{93}$Nb degrader positioned downstream of the target
further reduced the velocity to $\beta_L \sim$0.35. Depending on
whether the decay occurred in-flight between the target
and the degrader or after slowing down in the degrader, the
$\gamma$-rays exhibit different Doppler shifts. Consequently, the
$\gamma$-ray spectra contain two peaks for each transition. The
lifetime of the state can be inferred from relative intensities of
the peaks as a function of target/degrader separation using the
information on the ion velocity contained in the Doppler shift.

In the current studies the Doppler-shifted $\gamma$-rays were recorded
by the Segmented Germanium Array\ \cite{mue01} with two rings of 7 and
8 detectors at a laboratory angle of 30$^\circ$ and 140$^\circ$,
respectively. The data were recorded for target/degrader separations
of 0-, 200-, and 500-$\mu$m. In addition, a run without the degrader
was performed to measure the velocity of the reaction products
downstream from the target. This run also provided information on the
relative population of excited states from the reaction of
interest. 

A novel procedure of data analysis was developed specifically to
address experimental RDM information for states with lifetimes on the
order of a few ps; comparable to the time needed to cross the target
and/or degrader thickness. For such short lifetimes, procedures
discussed in Ref.\ \cite{Che06} for states with $\tau \sim$50 ps may
lead to systematic errors since the contribution of decays in the
target or degrader are significant and need to be accounted for. Thus
in the current studies lineshapes for transitions of interests were
first calculated as a function of the transition lifetime. Next the
lifetime was extracted from the comparison of these calculated
lineshapes to the experimental data using the least square fitting
method. The quality of calculated lineshapes for lifetimes yielding
the best agreement with the data is illustrated in Fig.\ \ref{fig1}.

While the details of the above analysis will be presented in a
separate paper it is worthwhile to stress a few aspects of the
procedure here. The parameters which define the line-shape for a given
target/degrader separation, besides the level lifetimes, are the
velocities of the nuclei of interest at the moment of the gamma
emission and the geometrical dimension and energy resolution of the Ge
detectors in use. In the current experiment the information on the
velocities of the incoming beam ions and the corresponding outgoing
reaction products is defined within 2\% and 6\% by the settings of the
A1900 and the S800 separators, respectively. The actual stopping
powers, which impact the line-shape calculations, are defined at the
intermediate energies by atomic processes and can be modeled quite
accurately. Only very small modifications were needed to reproduce the
measured velocities of the recoiling ions after they have passed the
target and the degrader.  The response of the SeGA array is understood
from the off-line source calibrations and Lorentz transformation from
the source to the laboratory reference frame.  The energy and angular
straggling of the reaction products as well as the energy resolution
of the gamma detectors are described by a single parameter which
enters as a width-parameter into Gaussian functions out of which the
line-shape is composed. In the present analysis four different
width-parameter values were used corresponding to decays occurring in
or after the target and for the observation of the gamma transitions
in the 30$^\circ$ or 140$^\circ$ Ge ring of the SeGA array.  These
values were selected to give a good representation of the width of the
two components of gamma-ray transitions observed in the spectra
including these measured with the plunger target only.  Thus having
fixed the parameters for the line-shape calculations the level
lifetimes were deduced from a fit of the calculated line-shapes to the
measured spectra; the only free parameters of the fit were the
lifetimes of interest and normalization factors to account for
different statistics accumulated at different target-degrader
distances and observation angles.

In the plunger experiments at intermediate energies the beam crossing
the target has enough energy to react in the degrader. It was measured
that in the current experiment 40\% of the observed excitations come
from the reaction on the degrader. This value of 40\% was used in all
fits for $^{64}$Ge and also for $^{62}$Zn.

In case of $^{62}$Zn the 2$^+_1$ lifetime was determined by taking
into account the 10\% feeding from the 2$_2^+$ state with the
3.8(6)~ps lifetime given in the literature\ \cite{jun00}.  The
extracted lifetime of 4.2(7)~ps is in excellent agreement with the
4.2(3)~ps lifetime of Ref.\ \cite{jun00}. Moreover, separate studies of the
impact of the unobserved feeding on the measured value indicate that
$\sim$90\% of the intensity of the $2^+_1$ decay in $^{62}$Zn comes
from fast feeding. This observation makes the knock-out reaction an
excellent tool for lifetime measurements.

In $^{64}$Ge a significant feeding of $\sim$30\% via the 677-keV
transition from the $2^+_2$ to the $2^+_1$ level was observed, see
Fig.\ \ref{fig1}.  Thus, the lifetime of the $2^+_2$ state was
fitted to the data shown in Fig.\ \ref{fig1} utilizing a single
exponential decay, while the corresponding fit for the $2^+_1$ state
was done taking into account the observed feeding.  The results
for the first measurements of the $2^+_1$ and $2^+_2$ state lifetimes
of the N=Z $^{64}$Ge are 3.3(5) and 8$^{+4}_{-2}$ ps, respectively.

From the measured lifetimes and data in Refs.\ \cite{sin96} and
\ \cite{jun00}, experimental observables were extracted and compared to
the Shell Model GXPF1A calculations as summarized in Tab.\
\ref{resSM}.  The GXPF1A interaction was recently
used\ \cite{ni56} to describe the low-lying states and the first
rotational band of $^{56}$Ni, by considering 16 valence particles in
the $pf$ model space ($^{64}$Ge is described by 16 valence
holes). Reference\ \cite{kan04} suggests that the contribution of the
$g_{9/2}$ orbital may be small in Zn and Ge, while the Nilsson diagram
suggests that the contribution of the $f_{7/2}$ orbital may be
important in this region.  The canonical effective charges of
$e_p=1.5\;e$ and $e_n=0.5\;e$ were used for the B($E2$)
calculations. It should be stressed that the agreement between
calculated and observed excitation energies for $^{64}$Ge is
better than 50~keV, while the transition rates are reproduced within
the experimental errors, except for the weak $2^+_2\rightarrow
0^+_1$ transition. A similar level of agreement is reached for
$^{62}$Zn. In both cases, the theory predicts a negative and positive
quadrupole moment for the $2^+_1$ and the $2^+_2$ states, respectively.

\begin{table}
\caption{Comparison between the experimental data and the
GXPF1A Shell Model calculations for $^{64}$Ge and $^{62}$Zn.}
\begin{ruledtabular}
\begin{tabular}{|c|c|c|c|c|}
nucleus &observable  & exp. & th. & unit\\
\hline\hline
$^{64}$Ge & E($2^+_1$) & 0.902 & 0.938 &MeV\\
$^{64}$Ge &E($0^+_2$) &       & 1.353 &MeV\\
$^{64}$Ge &E($2^+_2$) & 1.579 & 1.559 &MeV\\
$^{64}$Ge &E($4^+_1$) & 2.053 & 1.995 &MeV\\
\hline
$^{64}$Ge &B$(E2,2^+_1 \rightarrow 0^+_1$) & 410(60)&  406 & $e^2$fm$^4$\\
$^{64}$Ge &B$(E2,2^+_2 \rightarrow 2^+_1$) & 620(210)&  610 & $e^2$fm$^4$\\
$^{64}$Ge &B$(E2,2^+_2 \rightarrow 0^+_1$) & 1.5(5) &  14   & $e^2$fm$^4$\\
$^{64}$Ge &B$(E2,0^+_2 \rightarrow 2^+_1$) &        &  483  & $e^2$fm$^4$\\
$^{64}$Ge &B$(E2,0^+_2 \rightarrow 2^+_2$) &        &  12   & $e^2$fm$^4$\\
$^{64}$Ge &B$(E2,4^+_1 \rightarrow 2^+_1$) &        &  674  & $e^2$fm$^4$\\
$^{64}$Ge &B$(E2,4^+_1 \rightarrow 2^+_2$) &        &   9   & $e^2$fm$^4$\\
\hline
$^{64}$Ge &Q($2^+_1$)                     &        &-18.6   & $e$fm$^2$\\
$^{64}$Ge &Q($2^+_2$)                     &        &+18.5   & $e$fm$^2$\\

\hline\hline
$^{62}$Zn &E($2^+_1$) & 0.954 & 1.012 &MeV\\
$^{62}$Zn &E($2^+_2$) & 1.805 & 1.908 &MeV\\
\hline
$^{62}$Zn &B$(E2,2^+_1 \rightarrow 0^+_1$) & 250(18)&  295 & $e^2$fm$^4$\\
$^{62}$Zn &B$(E2,2^+_2 \rightarrow 2^+_1$) & 290(50)&  231 & $e^2$fm$^4$\\
$^{62}$Zn &B$(E2,2^+_2 \rightarrow 0^+_1$) & 4.5(7) &  11  & $e^2$fm$^4$\\
\hline
$^{62}$Zn &Q($2^+_1$)                     &        &-22.3   & $e$fm$^2$\\
$^{62}$Zn &Q($2^+_2$)                     &        &+13.8   & $e$fm$^2$\\
\end{tabular}
\end{ruledtabular}
\label{resSM}
\end{table}

The opposite quadrupole moments can be qualitatively explained by
large-amplitude collective dynamics. In the case of an anharmonic
vibrator Hamiltonian, which seems applicable based on the observed
excitation energy pattern and from microscopic calculations of Ref.\
\cite{kan04}, the most important anharmonicity is quartic
($~\alpha^{4}$) in the quadrupole coordinate ($\alpha$).  The
semi-microscopic estimates of various types of anharmonicity were
given in\ \cite{zel93} based on the soft quadrupole mode with low
quadrupole frequency, relatively close to the RPA instability.  Here
the quadrupole frequency is $\sim$1/2 of the pairing gap, i.e. not very
low.  In the limiting case of strong quartic anharmonicity, the
prediction\ \cite{vor83,vor85} is $R=E(4_1)/E(2_1)=2.09$. The perfect
case is $^{100}$Pd, where all states of the yrast band practically
coincide with predictions of strong quartic anharmonicity that is
characterized by O(5) symmetry.  In the case of $^{64}$Ge $R=2.13$,
 close to this limit are $^{70,72}$Ge with $R=2.07$.

The cubic anharmonicity, discussed first in Ref.\ \cite{bri65}, is
related to single-particle levels changing with deformation; this can
give a first order phase transition to static deformation. In the
limiting case of strong quartic anharmonicity, the cubic term is
typically small. This is analogous to the three-phonon vertex in
quantum electrodynamics that is strictly forbidden by the Furry
theorem (virtual contributions of electrons and positrons cancel
exactly).  For N=Z nuclei this would be the case for exact
particle-hole symmetry; in the BCS theory the effect is proportional
to the sum of contributions ($u^{2}-v^{2}$) which almost cancel, but
not exactly. It can be observed from a Nilsson diagram that for
$^{64}$Ge the lowest energy states which originate from the $g_{9/2}$
orbital go down for both signs of deformation, while the hole level
originating from the $f_{7/2}$ orbital goes up on the prolate side\
\cite{footnote}.  With a large amplitude of zero point quadrupole
motion, the nucleons probe all these shapes, and the cubic
anharmonicity is relatively important.  Moreover, it should be
stressed that the effect is enhanced by the coherent action of protons
and neutrons occupying the same shells.  The cubic anharmonic term
mixes the states with phonon numbers differing by one unit (its
contribution to energy comes only in the second order and therefore is
not large). For the $2^+_1$ and the $2^+_2$ states this mixing results
in quadrupole moments equal in magnitude and opposite in sign as given
in Tab.\ \ref{resSM}.

If the amplitude of zero point vibration is large and various
deformations are probed, there appears ``virtual rotation''\
\cite{vor85} based on slowly evolving dynamic deformation. This splits
the two-phonon states by $\frac{1}{2J}\times I(I+1)$, for $^{64}$Ge
$\frac{1}{2J}=31$ MeV. For good rotors the ``Alaga ratio'' of the
absolute value of the quadrupole moment in the lowest $2^+$ state to
the $\sqrt{\mbox{B}(E2)}$ from this state to the ground state is
2/7. Here we have instead 0.9, which means that the transition
probabilities are much weaker than it would be for a good rotor.

In summary, picosecond RDM lifetime measurements were performed using
a unique combination of state-of-the-art instruments and knock-out
reactions with rare isotope beams at the NSCL.  Studies of this
type hold the promise of reaching far from stability and providing
lifetime information for intermediate-spin excited states in a wide
range of nuclei. The absolute $E2$ transition rates measured here for
the $2^+_1$ and $2^+_2$ states in N=Z $^{64}$Ge are in excellent agreement
with state-of-the-art large-scale Shell Model
calculations. Theoretical analysis suggests that $^{64}$Ge is a
$\gamma$-soft anharmonic vibrator.

This work is supported by the US NSF under Grants No. PHY-0606007,
PHY-0555366 and MRI PHY-0619497 and also partly by the BMBF (Germany)
under Contract No. 06K167 and GSI, F.u.E. Contract No.  OK/JOL.  The
authors acknowledge computational resources provided by the MSU High
Performance Computing Center and by the Center of High Performance
Scientific Computing, at Central Michigan University.


\begin{thebibliography}{99}
\bibitem{Che06} A. Chester {\it et al.} Nucl. Instr.\& Meth. A{\bf 562}, 230
(2006).

\bibitem{mil01} P. Miller {\it et al.}, in:{\it Proc. of the
2001 Part. Acc. Conf.,} Chicago, IL, IEEE 01CH37268,
2001. p. 2557.

\bibitem{mor03} D.J.Morrissey {\it et. al}, Nucl. Instr.\& Meth.{\bf
B204}(2003) 90.

\bibitem{sto06} A. Stolz {\it et al.}, Diamond \& Related Materials
15, 807 (2006).

\bibitem{mue01} W.F.Mueller {\it et al.}, Nucl. Instr. \& Meth. {\bf
A466}, 492 (2001).

\bibitem{dew06} A. Dewald {\it et al.} in {\it GSI Scientific Report
2005}, 38 (2006).

\bibitem{baz03} D. Bazin {\it et al.}, Nucl. Instr. \& Meth. B {\bf
204}, 629 (2003).

\bibitem{kan04} K.~Kaneko, M.~Hasegawa, and T.~Mizusaki,
Phys. Rev. C{\bf 70}, 051301(R) (2004) and references therein.

\bibitem{fis99} S.M.~Fischer {\it et al.}, Phys. Rev. Lett. {\bf 84},
4064 (2000).

\bibitem{gad05} A. Gade {\it et al}., Phys. Rev. Lett. {\bf 95},
022502 (2005);  Phys. Rev. Lett. {\bf 96}, 189901(E) (2006).

\bibitem{coo06} J. M. Cook,T. Glasmacher, A. Gade,
Phys. Rev. C{\bf73}, (2006) 024315.

\bibitem{gx1a}
M. Honma, T. Otsuka, B.A. Brown and T. Mizusaki,
Eur. Phys. Jour. A {\bf 25} Suppl. 1, 499 (2005).

\bibitem{far03} E. Farnea {\it et al.} Phys. Lett. B {\bf 551}, 56 (2003).

\bibitem{sin96} B. Singh, Nuclear Data Sheets {\bf 78}, 395 (1996).


\bibitem{ni56}
M. Horoi, B.A. Brown, T. Otsuka, M. Honma and T. Mizusaki,
Phys. Rev. C {\bf 73}, 061305(R) (2006).

\bibitem{rg}
M. Hjorth-Jensen, T.T.S. Kuo and E. Osnes,
Phys. Rep. {\bf 261}, 125 (1995).


\bibitem{gx1}
M. Honma, T. Otsuka, B.A. Brown and T. Mizusaki,
Phys. Rev. C {\bf 65}, 061301(R) (2002);
Phys. Rev. C {\bf 69}, 034335 (2004).



\bibitem{sto05} A. Stolz {\it et al.}, Nucl. Instr. \& Meth. B {\bf
241}, 858 (2005).


\bibitem{you99} J. Yurkon {\it et al.}, Nucl. Instr. \& Meth. A {\bf
422}, 291 (1999).

\bibitem{jun00} H. Junde and B. Singh, Nucl. Data Sheets 91, 317
(2000).

\bibitem{Geant} S. Agostinelli {\it et al.}, Nucl. Instr. \& Meth. A
{\bf 506}, 250 (2003).

\bibitem{Lise} O.B. Tarasov and D. Bazin, Nucl. Phys. A{\bf 746}, 411c (2004).

\bibitem{zel93}  V.G. Zelevinsky, Int. J. Mod. Phys. E {\bf 2}, 273 (1993).


\bibitem{vor83} O.K. Vorov and V.G. Zelevinsky, Yad. Fiz. {\bf 37},
 1392 (1983); [Sov. J. Nucl. Phys. {\bf 37}, 830 (1983)].

\bibitem{vor85} O.K. Vorov and V.G. Zelevinsky, Nucl. Phys. A {\bf
439}, 207 (1985).

\bibitem{bri65} D.M. Brink, A.F.R. De Toledo Piza, and A.K. Kerman,
Phys. Lett.  {\bf 19}, 413 (1965).

\bibitem{footnote} For that reason the $f_{7/2}$ state has to
be included in the Shell-Model space.


\end{thebibliography}
\end{document}